\documentclass[10pt,twocolumn,letterpaper]{article}

\usepackage{cvpr}              

\usepackage{graphicx}
\usepackage{amsmath}
\usepackage{amssymb}
\usepackage{booktabs}
\usepackage[pagebackref,breaklinks,colorlinks]{hyperref}

\usepackage[capitalize]{cleveref}
\crefname{section}{Sec.}{Secs.}
\Crefname{section}{Section}{Sections}
\Crefname{table}{Table}{Tables}
\crefname{table}{Tab.}{Tabs.}

\begin{document}

\title{\ Multi-Stage Graph Learning for fMRI Analysis to Diagnose Neuro-Developmental Disorders}
\author{
    Wenjing Gao\thanks{Guangzhou Women and Children's Medical Center }, 
    Yuanyuan Yang\thanks{South China Normal University },
    Jianrui Wei\thanks{Guangzhou Women and Children's Medical Center},
    Xuntao Yin\thanks{Guangzhou Women and Children's Medical Center},
    Xinhan Di\thanks{Giant Network AI Lab},
}

\maketitle
\begin{abstract}
The insufficient supervision limit the performance of the deep supervised models for brain disease diagnosis. It is important to develop a learning framework that can capture more information in limited data and insufficient supervision. To address these issues at some extend, we propose a multi-stage graph learning framework which incorporates 1) pretrain stage : self-supervised graph learning on insufficient supervision of the fmri data 2) fine-tune stage : supervised graph learning for brain disorder diagnosis. Experiment results on three datasets, Autism Brain Imaging Data Exchange ABIDE I, ABIDE II and ADHD with AAL1, demonstrating the superiority and generalizability of the proposed framework compared to the state of art of models. (ranging from 0.7330 to 0.9321,0.7209 to 0.9021,0.6338 to 0.6699)
\end{abstract}

\section{Introduction}
\label{sec:intro}
Autism spectrum disorder (ASD) is a developmental disorder characterized by significant social, communication, and behavioral challenges\cite{hirota2023autism,bouzy2023transidentities}. Recent data from the Centers for Disease Control and Prevention (\url{https://data.cdc.gov/}) indicate that one in 36 children is diagnosed with autism. However, current diagnostic methods heavily rely on traditional behavioral assessments, which are subjective and can often result in missed early symptoms and misdiagnoses. Consequently, there is a growing interest in identifying objective biomarkers for early diagnosis and timely intervention in ASD treatment within the fields of psychiatry and neuroscience\cite{lord2020autism}.

Resting-state functional MRI (rs-fMRI) is a technique that measures blood-oxygen-level-dependent (BOLD) signals in the brain without requiring the subject to perform specific tasks. This technique has been widely employed to identify potential neuroimaging biomarkers for various psychiatric disorders\cite{millevert2023resting}. Functional connectivity networks (FCNs) are constructed to represent each subject, with each element indicating the pairwise relationships between brain regions of interest (ROIs). FCNs help us understand brain organization patterns and diagnose psychiatric disorders such as ASD\cite{lepping2022visuomotor}, attention-deficit/hyperactivity disorder (ADHD)\cite{guo2020shared} depressive disorders\cite{yan2019reduced}, bipolar disorder\cite{hu2023aberrant}, and anxiety disorders \cite{rezaei2023machine}. However, previous studies typically extract handcrafted network features (such as node degree and clustering coefficient) from FCNs and then use these features in prediction models for ASD diagnosis. This approach limits diagnostic accuracy and heavily depends on expert knowledge.

\section{Related work}
\subsection{Graph Network for fMRI}
As graph convolutional network (GCN) can well handle irregular graph structures with the graph convolution to propagate features of adjacent vertexes\cite{scarselli2008graph}, GCN based models have achieved greater successes in learning FCN feature representation (namely FCN embedding), and consequently have improved the diagnostic performance for brain diseases 
\cite{ktena2018metric,qin2022using,chu2022multi}. A systematic study of GCN-based brain network analysis with models, examples and out-of-box Python package has been presented\cite{cui2022braingb}. For instance, a Siamese GCN model was proposed to learn a graph similarity metric between FCNs for distinguishing autism patients from healthy controls\cite{ktena2018metric}. Another study investigated a graph embedding learning (GEL) model for diagnosing major depressive disorder by learning FCN embeddings via GCN and then using a fully connected layer activated by a soft-max function for final classification\cite{qin2022using}. However, the above GNN network is designed for a single training stage and produce low accuracy, therefore, we proposed the multi-stage graph for fMRI analysis.

\subsection{Multi-Stage Graph Learning}
Self-supervised learning (SSL) has emerged as a promising paradigm that leverages the inherent structure of the data to provide supervision, particularly useful in scenarios with limited labeled data. Despite its success in computer vision (CV) and natural language processing (NLP)\cite{chen2024context,eckart2021self,akbari2021vatt}. However, SSL are not wildely applied to the fMRI analysis, therefore, we proposed the Multi-Stage graph learning to the analysis of fMRI.

\section{Methods}
As illustrated in Fig. 1, Multi-Stage Graph comprises two primary modules: a self-supervised learning with a contrastive learning adversarial strategy hypergraph (Multi-Stage Graph pre-training) and a graph classification model (fine-tuning). 
The Multi-Stage Graph pre-training involves creating an original hypergraph and its edge-dropped version for contrastive learning and feature distillation. During contrastive learning, the model distinguishes between these two versions, using a trainable Bernoulli mask to randomly drop edges, thus encouraging the learning of invariant features. The distilled features, robust to graph variations, are crucial for disease diagnosis. The remaining edge weights after training can be analyzed as potential biomarkers. 
In the fine-tuning stage, these distilled features are classified using a linear classifier trained to differentiate between ASD and NC subjects based on connectivity patterns and functional relationships within the brain. 

\subsection{Pre-training Stage}
In this stage, to train the feature extractor and the parameters in the graph augmentation process, we use a loss function that forces the two feature vectors to be close if they are from the same graph and far away if they are from different graphs. In this way, the trained feature extractor can keep the most important information while removing excessive information in the graph. Since the training proceeds in a batch-wise manner, such a loss is also imposed batch-wisely. Specifically, the loss is defined by the infoMax principle (R Devon Hjelm et al., 2018) which we aim to maximize:
\begin{align}
L(\mathcal{F},e,N_{g})=\frac{1}{|N_{g}|}\sum_{g\in N_{g}}l o g\frac{\exp{(s i m(\mathcal{F}(g),\mathcal{F}(\mathcal{G}))))}}{\sum_{\tilde{g}\in N_{g}}\exp{(s i m(\mathcal{F}(g),\mathcal{F}(\tilde{g}^{\prime}))}}
\end{align}

where $N_g$ is the set of graphs in a batch, $\left | N_g \right |$  is its cardinality, sim is the similarity metric, and we use the cosine of the angle between two input vectors: 

\begin{align}
\operatorname{sim}\left(\mathcal{F}(g), \mathcal{F}(\widetilde{g})=\mathcal{F}(\widetilde{g})^{T} \mathcal{F}(g) /\|\mathcal{F}(\widetilde{g})\|\|\mathcal{F}(g)\|\right.
\end{align}
Maximizing $L\left(\mathcal{F}, e_{u v}, N_{g}\right)$ could be easily achieved by retaining all the edges. To force more edges to be dropped in this context, we need a regularization term to facilitate edge dropping. Such a regularization term is designed to be the mean of all the $e_{u v} \text { 's }$ and minimize $L\left(e_{u v}, N_{g}\right)$. 
\begin{align}
L\left(e, N_{g}\right)=\frac{1}{\left|N_{g}\right|} \sum_{g \in N_{g}} e(g)
\end{align}
Finally, the total objective function is defined as:
\begin{align}
L_{\text {total }}=L\left(\mathcal{F}, e, N_{g}\right)-L\left(e, N_{g}\right)
\end{align}

where the optimal $e$ and $F$ are obtained through gradient descent/ascent of the corresponding parameters.
\begin{figure*}{}
\centering
\includegraphics[width=1\textwidth]{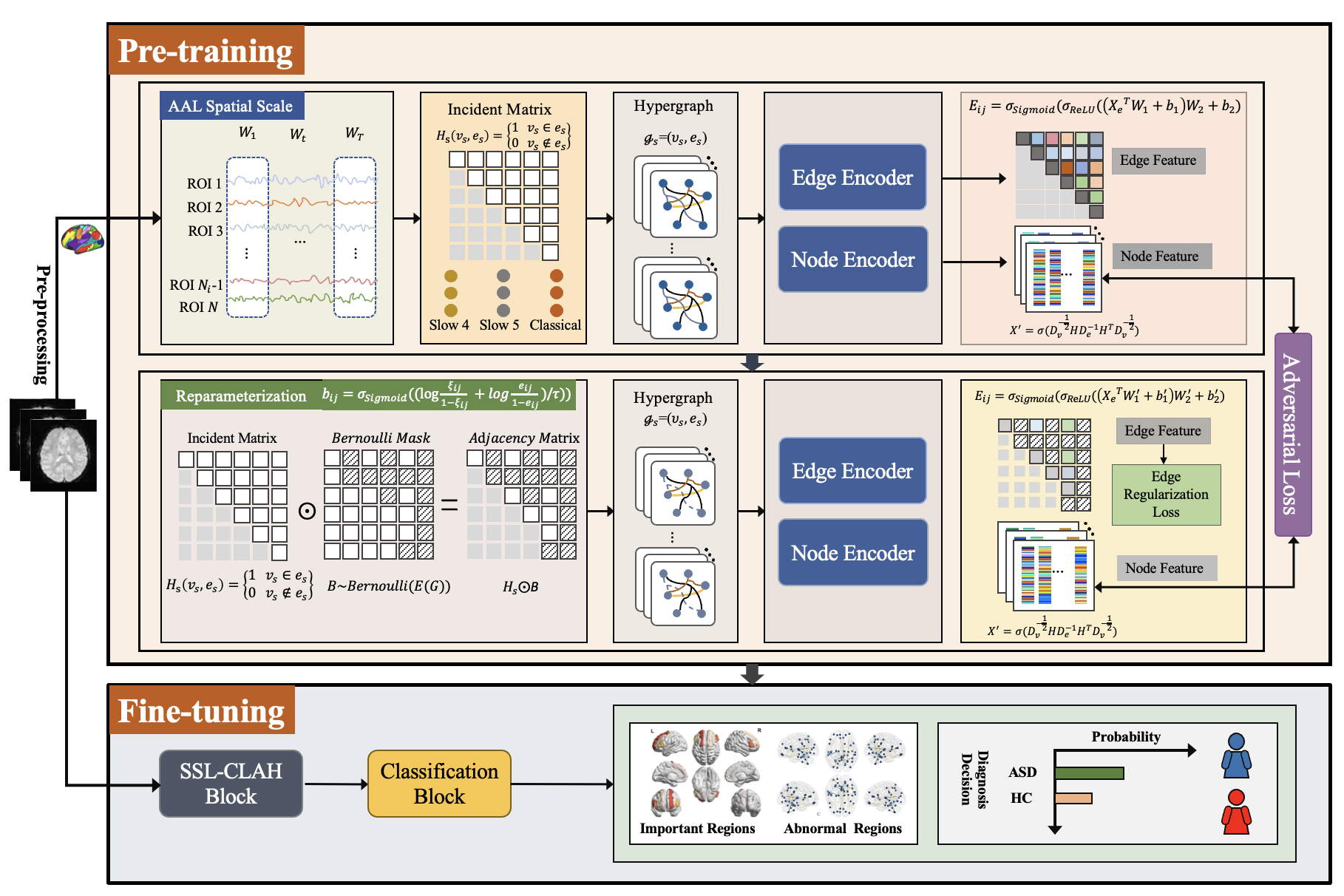}
\caption{Illustration of the proposed Multi-Stage Graph. The model can be divided into a hypergraph autoencoder (pre-training) and a hypergraph classification model (fine-tuning). (1) We propose a hypergraph autoencoder which consists of three stages: edge embedding learning and graph embedding learning. Multi-Stage Graph performs contrastive learning on features extracted from the original hypergraph and its edge-dropped version based on a Bernoulli mask, enabling the learned features to be independent of the class labels and genuinely represent an embedding of the graphs in Euclidean space. (2) The pre-trained hypergraph autoencoder and fMRI BOLD signals are then combined to obtain the final graph embedding and diagnose results under the supervision of the class labels.}
\label{fig1}
\end{figure*}

\subsection{Fine-tuning stage}
In this stage, with the signal representation learning module and the pretrained topology-aware encoder, the node embeddings learned from both the graph structure and BOLD signals can be utilized in the downstream graph classification task. First, we fuse the two sets of node embeddings using a simple summation. Then, a 1D convolutional layer is employed to convert the fused node embeddings into the final graph representation. Finally, a multi-layer perceptron (MLP) is used to predict the class of the input graphs. The hypergraph classification model is fine-tuned using cross-entropy loss to optimize the classification performance.

\begin{align}
\mathcal{L}_{C E}=\min \frac{1}{\mathcal{N}} \sum_{p}^{\mathcal{N}}-\left[y_{p} \cdot \log \left(\hat{y}_{p}\right)+\left(1-y_{p}\right) \cdot \log \left(1-\hat{y}_{p}\right)\right]
\end{align}

\section{Experimental Evaluation}
\subsection{Experimental setup}
Our model is implemented in PyTorch. All of the algorithms described in this paper can be executed on a single GPU. The experiments are accelerated by two servers with two NVIDIA V-100 GPUs.  The implementation details are as follows: The learning rate is set to 0.001. The hypergraph embedded dimension $d$ is set to 32. The batch size is 32. The temperature $\tau \mathrm{i}$ is set to 1. When applying the proposed model to a new fMRI dataset, it is recommended to search the batch size in \{8, 16, 32, 64\}.

\begin{table*}[]
\setlength{\abovecaptionskip}{0cm} 
\setlength{\belowcaptionskip}{-0.2cm}
\vspace{0.05cm} 
\caption{Benchmark for ABIDE I , ABIDE II and  ADHD.
The comparison results with DGCN,GATE,A-GCL and our proposed Muti-Stage Graph.}
\centering
\setlength{\tabcolsep}{15mm}
\begin{tabular}{|l|l|l|l|}
\hline
Dataset                                        & Methods           & ACC    & AUC    \\ \hline
\multicolumn{1}{|c|}{{\centering ABIDE I}} & DGCN              & 0.7330 & 0.7415 \\ \cline{2-4} 
\multicolumn{1}{|c|}{}                         & GATE              & 0.7352 & 0.7560 \\ \cline{2-4} 
\multicolumn{1}{|c|}{}                         & A-GCL             & 0.8065 & 0.8142 \\ \cline{2-4} 
\multicolumn{1}{|c|}{}                         & Multi-Stage Graph & 0.9321 & 0.9318 \\ \hline
{ABIDE II}                      & DGCN              & 0.7258 & 0.7327 \\ \cline{2-4} 
                                               & GATE              & 0.7209 & 0.7406 \\ \cline{2-4} 
                                               & A-GCL             & 0.7988 & 0.8004 \\ \cline{2-4} 
                                               & Multi-Stage Graph & 0.9021 & 0.9408 \\ \hline
{ADHD}                          & DGCN              & 0.6338 & 0.6504 \\ \cline{2-4} 
                                               & GATE              & 0.6526 & 0.6572 \\ \cline{2-4} 
                                               & A-GCL             & 0.7092 & 0.7112 \\ \cline{2-4} 
                                               & Multi-Stage Graph & 0.6699 & 0.6449 \\ \hline
\end{tabular}
\end{table*}

\subsection{Evaluation}
We compared the performance of different methods on the ABIDE I, ABIDE II, and ADHD datasets, including DGCN, GATE, A-GCL, and our proposed Multi-Stage Graph. Table 1 shows the accuracy (ACC) and area under the curve (AUC) for each method on these datasets.
On the ABIDE I dataset, the ACCs of the DGCN, GATE, and A-GCL methods were 0.7330, 0.7352, and 0.8065, respectively, with corresponding AUCs of 0.7415, 0.7560, and 0.8142. Our Multi-Stage Graph method significantly outperformed these methods, achieving an ACC of 0.9321 and an AUC of 0.9318.
On the ABIDE II dataset, the ACCs of the DGCN, GATE, and A-GCL methods were 0.7258, 0.7209, and 0.7988, respectively, with corresponding AUCs of 0.7327, 0.7406, and 0.8004. Our Multi-Stage Graph method also performed excellently, with an ACC of 0.9021 and an AUC of 0.9408.
On the ADHD dataset, the ACCs of the DGCN, GATE, and A-GCL methods were 0.6338, 0.6526, and 0.7092, respectively, with corresponding AUCs of 0.6504, 0.6572, and 0.7112. Although the ACC of our Multi-Stage Graph method was 0.6699, slightly lower than A-GCL, its AUC was 0.6449, demonstrating stable performance compared to other methods. The proposed Multi-Stage Graph method significantly outperformed other methods on the ABIDE I and ABIDE II datasets and showed competitive performance on the ADHD dataset, validating the effectiveness and robustness of our method in brain network analysis.

\begin{figure*}{}
\centering
\includegraphics[width=1\textwidth]{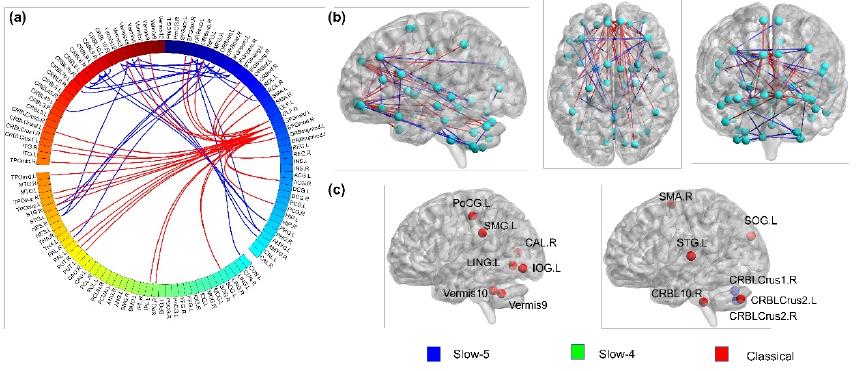}
\caption{Model interpretability on ABIDE I. (a) All 66 selected functional connection features shown in the circus plot. (b) 66 functional connections on the cortical surface. The red and blue lines in (a) and (b) represent the increased and decreased strength functional connections in ASD compared to TD, respectively. (c) Increased node features in ASD compared to TD. (d) Decreased node features in ASD compared to TD. In (c) and (d), the red, green, and blue bubbles represent the Slow-5, Slow-4, and classical ALFF features, respectively. }
\label{fig2}
\end{figure*}

\subsection{Ablations Study for ABIDE I , ABIDE II and ADHD.}
Table 2 presents the ablation study results for the ABIDE I, ABIDE II, and ADHD datasets, comparing the performance of the One-Stage-Graph and Multi-Stage-Graph methods. The results are reported in terms of accuracy (ACC) and area under the curve (AUC). For the ABIDE I dataset, the One-Stage-Graph method achieved an ACC of 0.8510 and an AUC of 0.8509. In contrast, the Multi-Stage-Graph method significantly improved the performance, with an ACC of 0.9321 and an AUC of 0.9318. This demonstrates the effectiveness of the multi-stage approach in capturing complex interactions in brain networks. Similarly, on the ABIDE II dataset, the One-Stage-Graph method obtained an ACC of 0.7255 and an AUC of 0.7281, whereas the Multi-Stage-Graph method substantially outperformed it with an ACC of 0.9021 and an AUC of 0.9408. This substantial improvement underscores the robustness of the multi-stage strategy in handling diverse and challenging datasets. In the case of the ADHD dataset, the One-Stage-Graph method achieved an ACC of 0.6409 and an AUC of 0.6416. The Multi-Stage-Graph method again showed better performance with an ACC of 0.6699 and an AUC of 0.6449. Although the improvement is relatively smaller compared to the ABIDE datasets, it still indicates the advantage of the multi-stage method in extracting meaningful features even in more difficult classification tasks. Overall, the Multi-Stage-Graph method consistently outperformed the One-Stage-Graph method across all datasets, validating the proposed multi-stage framework's superiority in brain network analysis.

\begin{table*}[]
\setlength{\abovecaptionskip}{0cm} 
\setlength{\belowcaptionskip}{-0.2cm}
\vspace{0.05cm} 
\caption{\raggedright Ablations Study for ABIDE I , ABIDE II and  ADHD.}
\setlength{\tabcolsep}{15mm}
\begin{tabular}{|l|l|l|l|}
\hline
Datasets                  & Methods           & ACC    & AUC    \\ \hline
{ABIDE I}  & One-Stage-Graph   & 0.8510 & 0.8509 \\ \cline{2-4} 
                          & Multi-Stage-Graph & 0.9321 & 0.9318 \\ \hline
{ABIDE II} & One-Stage-Graph   & 0.7255 & 0.7281 \\ \cline{2-4} 
                          & Multi-Stage-Graph & 0.9021 & 0.9408 \\ \hline
{ADHD}     & One-Stage-Graph   & 0.6409 & 0.6416 \\ \cline{2-4} 
                          & Multi-Stage-Graph & 0.6699 & 0.6449 \\ \hline
\end{tabular}
\end{table*}

\subsection{Model Interpretability}
The important brain connections. For edge features, we performed a transpose averaging operation on the gradient of edge features in the ASD group to create a saliency map with symmetric edges. Positive gradients indicate increased edge features in ASD compared to TD, while negative gradients indicate decreased features. We selected the top 66 connections ($top 1\%$) with the highest absolute values from a total of $6670$ ($0.5 \times 116 \times 115$) connections. These are shown in Figure. 2(a). Figure. 2(b) shows the distribution of these connections on the cortical surface. Red lines highlight connections with increased strength in ASD, while blue lines highlight those with decreased strength. In the ABIDE I dataset, the positive connections mainly link regions such as the SFG, SFGmed, ORBsupmed, and OLF, with the parietal cortex, cerebellum, and thalamus. Negative connections primarily link the orbitofrontal cortex and MFG with the cerebellum, temporal cortex, and limbic system (PHG and AMYG). In the ABIDE II dataset, the connections are more dispersed, with most showing increased strength (Figure.3).

For ALFF features, using the node saliency map of the ASD group defined in\cite{heinsfeld2018identification}, we identified the top $15$ nodes with the highest absolute values from a total of $348$ ($116 \times 3$) nodes. Positive gradients in the node saliency map represent increased node features in ASD, while negative gradients represent decreased features. Figure.2 show the cortical surface locations of the increased(Left) and decreased node features(Right) in ASD compared to TD, respectively. We further analyze the significant brain regions associated with ASD using Multi-Stage Graph. The importance of a brain region is quantified by the summed functional connectivity’s (FCs) of a node in the edge-dropped graph. For each dataset, we calculate this importance score for each brain region, sort the scores, and select the top $15$ most important regions. The results are visualized using BrainNet-Viewer (\url{www.nitrc.org}) and displayed in Fig. 5. As shown in Fig. 5, for the ABIDE I dataset, the most important regions for ASD classification include the cerebellum, bilateral PHG.R, PCUN.R, bilateral CUN, ACG.L, MFG.R, PCL.L, SFGmed.R, and ORBsup.R. For the ABIDE II dataset, the most important regions are the cerebellum, bilateral MFG, bilateral CUN, bilateral PCUN, SOG.R, SFGdor.L, ACG.R, bilateral PoCG, DCG.R, and SFGmed.R. Notably, some regions such as SFGmed, CUN, PCUN, and ACG are identified in both datasets and have been reported in previous studies\cite{monk2009abnormalities,chan2011abnormalities,retico2016effect,fan2021developmental}. SFGmed.L and SFGmed.R are the two regions with the highest node degree and the connections are all the positive connection. The med SFG is involved in higher-order social cognitive processes such as understanding others' intentions and emotions.

\textbf{The interpretation of the model.} We found that ASD patients showed decreased FC in the key nodes of the default mode network (DMN)\cite{doucet2019evaluation}, such as the medial prefrontal cortex/anterior cingulate cortex (mPFC/ACC) and precuneus in both datasets, and the bilateral parahippocampal gyri in ABIDE I. The DMN is involved in social cognition processing, which has been considered to play a central role in the physiopathology of ASD\cite{padmanabhan2017default,chen2021changes,curtin2022altered,jia2022aberrant}. Decreased nodal centralities\cite{chen2021changes}, local and long-range connectivity\cite{leung2020resting}in these regions during resting state, have been reported in adolescents and adults with ASD. In line with the resting-state results, previous structural and task-related functional MR studies also found that ASD patients showed morphologic abnormalities in the mPFC/ACC\cite{guo2024systematic},precuneus\cite{kitamura2021association} , parahippocampal gyrus\cite{monkul2003structural}, and less activation in the mPFC on self-relevant social reward task\cite{sumiya2020attenuated}. Therefore, our model supported the edge features of the DMN as a useful biomarker for the diagnosis of ASD. 

\begin{figure*}{}
\centering
\includegraphics[width=1\textwidth]{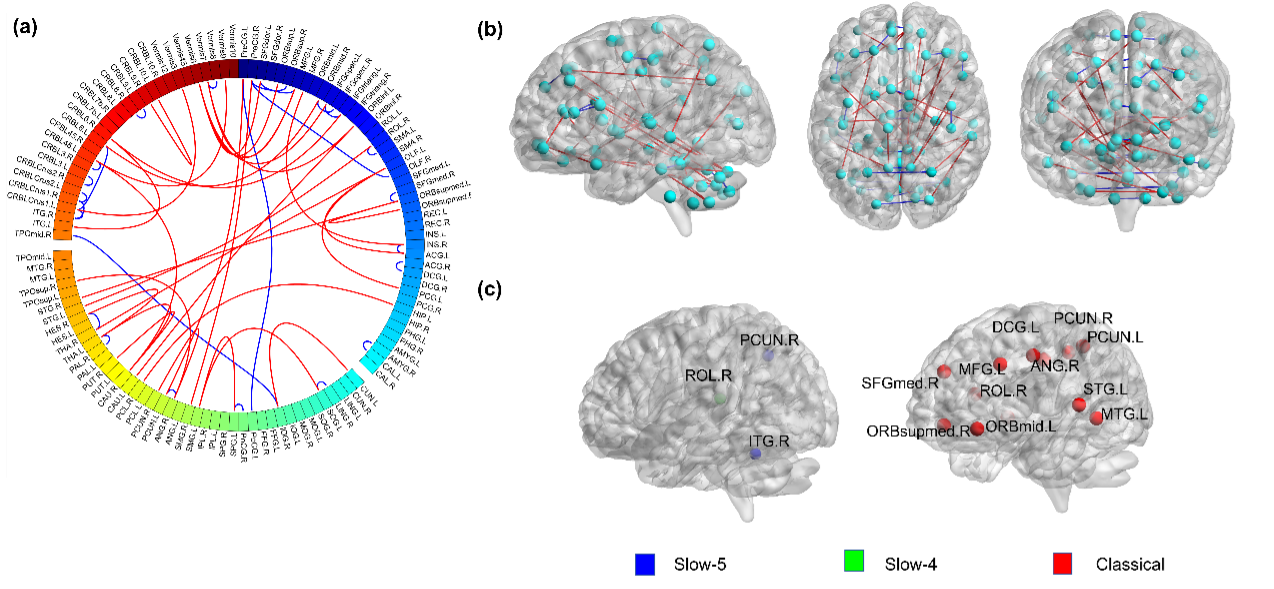}
\caption{Model interpretability on ABIDE II. (a) All 66 selected functional connection features shown in the circus plot. (b) 66 functional connections on the cortical surface. The red and blue lines in (a) and (b) represent the increased and decreased strength functional connections in ASD compared to TD, respectively. (c) Increased node features in ASD compared to TD. (d) Decreased node features in ASD compared to TD. In (c), the red, green, and blue bubbles represent the Slow-5, Slow-4, and classical ALFF features, respectively.}
\label{fig2}
\end{figure*}

\begin{figure}[htbp]
\centering
\includegraphics[width=1\columnwidth]{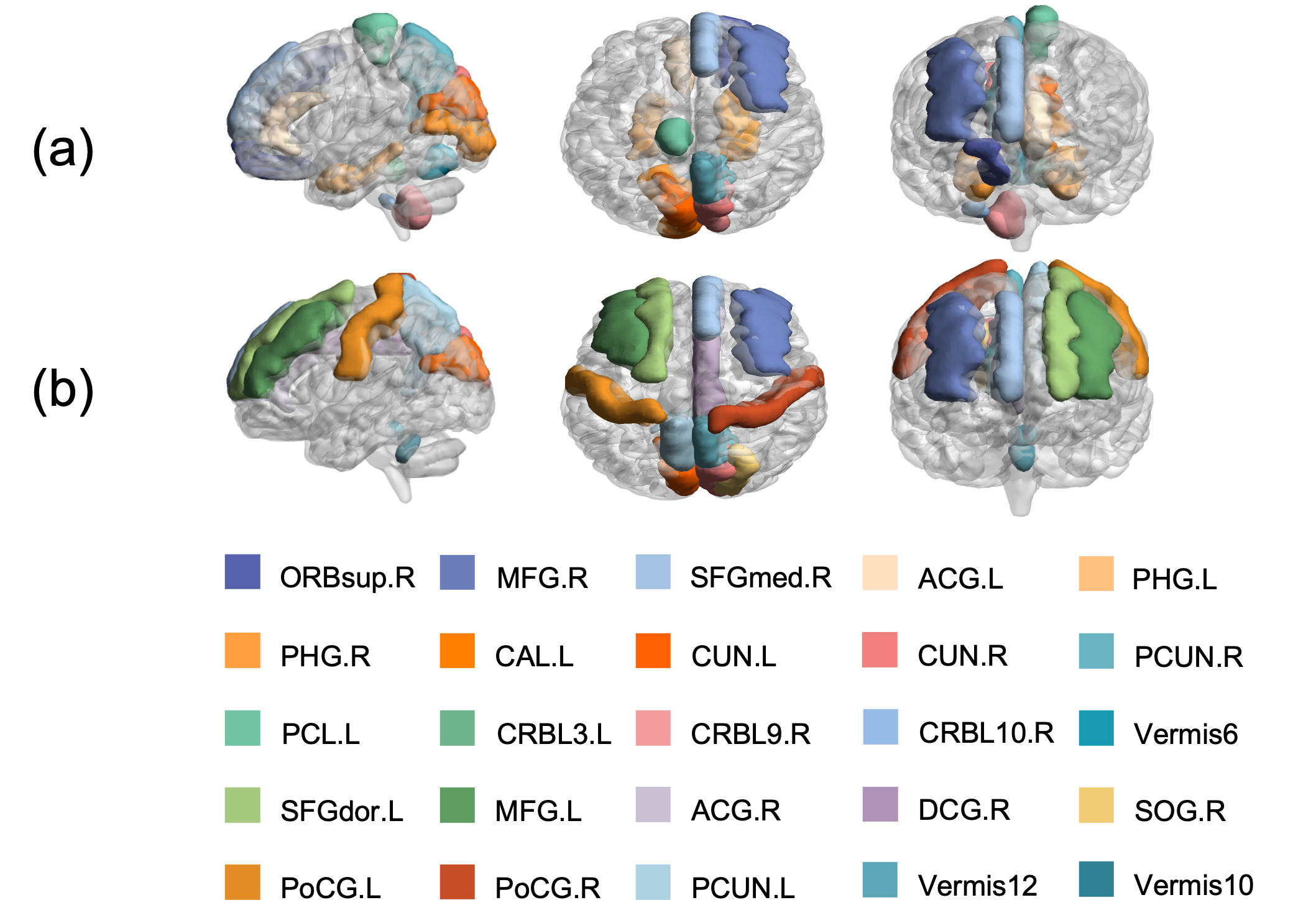}
\caption{Qualitative results of heatmaps generated by different baselines (the red mask represents lesion area, dark blue arrows represent various noise).}
\label{fig8}
\end{figure}

Our results also identified aberrant edge feature in the cuneus/calcarine fissure and dorsolateral prefrontal cortex (DLPFC) in ASD compared to TD group in both datasets. Meta-analysis had revealed that autistic people showed less prefrontal activity during perception tasks and greater recruitment of extrastriate cortex during visual processing\cite{jassim2021meta}. The cuneus is implicated in visual processing\cite{palejwala2021anatomy} and retrieval of autobiographical memories\cite{spreng2009common}. Connecting with other regions such as the orbitofrontal cortex, thalamus, and the neocortex (including posterior temporal, parietal, and occipital areas), the DLPFC is involved in goal-directed thought and action (Cole and Schneider 2007), and is linked to the non-social abnormality of ASD\cite{jassim2021meta}, such as restricted and repetitive behaviors, or reducing self-referential thoughts. Consequently, the DLPFC has been used as a target region of transcranial current\cite{qiu2021transcranial} or magnetic\cite{enticott2014double} stimulation for the treatment of ASD. The cerebellum is also a key node in motor learning, language comprehension, social skill, visual-spatial performance and memory functions\cite{sydnor2022structure}, which overlap with the aberrant domains observed among individuals with ASD. It has been pointed out that cerebellar areas have reduced structural or functional connectivity to the dorsolateral and medial prefrontal cortex\cite{kelly2021cerebellar}, which is related to the social deficits and increased repetitive behaviors in ASD mouse models and individuals with ASD\cite{kelly2020regulation}. Both of the functional connectivity and ALFF features of the cerebellum contribute to the diagnosing model, indicating that the cerebellar function is more vulnerable in ASD. In fact, ASD-associated genes were preferentially expressed in the cerebellum during the fetal stage\cite{hughes2023genetic}, and models that include cerebellum can significantly improve the accuracy.
Our results showed that node and edge features in the resting brain together could provide the good classification results in two ASD datasets. Aberrancies in dynamic functional interactions mainly represent in the medial and lateral prefrontal cortices, the posterior temporal and occipital areas, and the cerebellum. The functional organization at rest is expected to possess the intrinsic capability of supporting diverse cognitive processes, which may reveal the neural mechanism of ASD, and be used as the diagnostic and therapeutic indicators. 

\section{Discussion}

In this paper, we introduced a Multi-Stage Graph framework to extract FCN and ALFF embeddings, where hyperedge-aware self-supervised learning with a contrastive learning adversarial strategy in the pretraining stage was developed to capture complex information in brain network. In the funetuning stage, The graph embeddings learned from pretraining stage can be fine tuned for the downstream graph classification task. Finally, experimental results on the ABIDE demonstrated the superiority of our method over several comparable ones for ASD/HC classification. Furthermore, our method is interpretable and we successfully discovered ASD-relevant ROIs that have been previously reported in the literature. We are going to improve performance of three datasets (ABIDE I ,ABIDE II and ADHD) further with the support of the brain foundation model in the future.
{\small
\bibliographystyle{ieee_fullname}
\bibliography{egbib}
}

\end{document}